\documentclass[12pt,a4paper]{article}
\usepackage{epsfig}
\usepackage{amstex}

\newcounter{statement}

\newenvironment{statement*}[4]
  {\par\noindent#1#2 #4\unskip: #3}{\par\vspace{2mm}}
\newenvironment{Def}
  {\begin{statement*}{\sl}{Definition}{\rm}}{\end{statement*}}
\newenvironment{Prop}
  {\begin{statement*}{\sl}{Proposition}{\rm}}{\end{statement*}}
\newenvironment{Prf}
  {\begin{statement*}{\sl}{Proof}{\rm}}{\end{statement*}}
\newenvironment{Rem}
  {\begin{statement*}{\sl}{Remark}{\rm}}{\end{statement*}}

\begin{document}

\noindent
\rightline{MPI-PhT/97-83}
\rightline{Dec 1st, 1997}
\vskip1.0cm
\begin{center}
{\large \bf Remarks concerning an Entropy-like Quantity $S_q$\\
\vskip1mm
 especially for Quantum Systems with a  Particle Picture;\\
\vskip1mm
 Model with pure Point Spectra}
\end{center}
\vskip0.4cm

\begin{center}
S. Schlieder
\end{center}

\begin{center}
{\it Max-Planck-Institut f\"ur Physik \\
(Werner-Heisenberg-Institut) \\
F\"ohringer Ring 6, D-80805 Munich, Germany} \\
\end{center}

\vspace{4.0cm}

\begin{abstract}
For quantum mechanical systems an entropy-like quantity $S_q$ is
defined. $S_q$ can differ from the usually defined entropy $S$
and $S_q$ may increase with time for an isolated system. The
essential condition for the difference between $S$ and $S_q$ is
the assumption that the set {\bf A} of observables which can be
represented by a measurement is a proper subset of the set of selfadjoint
operators. The underlying idea is made visible in the case of particle systems
with non-trivial scattering. The model-character of the reasoning comes
from the fact that continuous spectra are replaced by point-spectra.
So it seems evident, that no direct connection exists between $S_q$ and the
Sinai-Kolmogorov-Entropy at least in this model with pure
point-spectra.
\end{abstract}

\newpage

\section{Introduction}
\setcounter{equation}{0}

Thermodynamics and the physical quantities appearing in it got
a deeper foundation by that branch of theoretical physics, which
one calls ``Statistical Mechanics'' and which is connected with the
work of Boltzmann and Gibbs. Using statistical methods the concepts
of atomism of matter were introduced into the framework of the older
thermodynamics. This way also the thermodynamic quantity ``Entropy''
got a new interpretation. Entropy from that time on expressed the
deficit of knowledge about the exact details of the atomistic picture of
a physical system -- for instance for a gas the lack of knowledge concerning
the position or the momentum of each particle as an individuum in the
ensemble representing this gas.

This statistical approach is then continued if one proceeds from classical
systems to quantum systems. Since one cannot characterize the state of the
system by using position and momentum of every particle at the same time,
one has to go over to the quantum state of the entire system into which for
instance -- in the case of a sufficient dynamical separation of each particle
from the other ones -- the quantum state of each particle is an ingredient.
The lack of knowledge will be described by a set of probabilities $\{w_i\}$
for the possible quantum states $\{\varphi_i\}$ respectively and one defines the
entropy $S$ by
\begin{equation}
S = -\sum w_i \ln w_i, \quad \sum w_i = 1
\end{equation}
If one regards  entropy in this way, defined as a function of time
then an irritating feature may arise in the cases were the $\{\varphi_i\}$
don't describe stationary states. Let $\{\varphi_i\}$ be an orthonormal basis
in the Hilbertspace ${\bf H}$ with the interpretation
$\{\varphi_i = \varphi_i(t_0)\}$ and $\{\varphi_i(t) = U(t,t_0) \varphi_i(t_0)\}$
describing the unitary time development of the Schr\"odinger states
$\{\varphi_i(t)\}$. The entropy itself however remains constant, since the
orthonormal basis $\{\varphi_i(0)\}$ transforms under the time development
into the one parameter set $\{\varphi_i(t)\}$ of orthonormal bases; this
means that the mixture $\{w_i\varphi_i(0)\}$ transforms into the
mixtures $\{w_i\varphi_i(t)\}$; so $S(t) = S(0)$ and the entropy
remains constant.

In one respect this feature is not a new experience: In the classical cases,
one defines entropy at a time {t} for instance at $t$= 0 usually by
the support $\Pi(0)$ of the probability measure at $t$ = 0 in the phase
space. Generically the support $\Pi(0)$ changes with time to supports
$\Pi(t)$; however as a consequence of one of the famous Louiville-Theorems,
the volume of $\Pi$ stays constant and thereby also the entropy connected
with this volumen remains also constant.

On the other side entropy -- as a quantitative expression for the lack of
knowledge about the system -- should generically increase with time.
Also one can learn in thermodynamics that the entropy of an isolated
system should increase with time except in the special cases, where it
has reached its internal equilibrium.

These controversal features are very well known  since a long time and
as a consequence new aspects were brought into the game to overcome
these difficulties. So one introduced besides the fine grain picture
in phase space a coarse grain picture by performing local smoothing of the
probability measure. Another approach to these problems is possible since
Kolmogorov's fundamental work. One uses an adequate decomposition of the
phase space to define a certain kind of entropy - conditioned by the
knowledge of the past (the so called ``Kolmogorov-Sinai-Entropy'').
This may increase in time with a specific rate for certain dynamical systems.

The following considerations are restricted to quantum-mechanical systems
and should lead to the definition of an entropy-like quantity $S_q$.
Usually the entropy-concept for quantum-mechanical systems is based
on the uncertainty in the knowledge about the quantum
states actually involved. Typically one expresses this fact with the use of mixtures of
pure states.

In contrast to the procedure described defining the entropy for a
quantum mechanical system we intend to take into account also the
question whether all virtual informations contained in a quantum state
can be made into real information by experiments. The generalized answer
to this question leads to the definition of an entropy-like quantity
$S_q$ in section 3. A general feature of $S_q$ is, that it can be
$\not = 0$ also for pure states and that $S_q$ can also grow with time
for one and the same quantum state. To restrict the discussion to this
main point, we specialize the discussion to states which are pure before 
measurements are performed. The increase of $S_q$ with time is a common
feature with the increase of the Kolmogorov-Sinai-entropy.
However the reasons for the increase are quite different; one sees this
immediately regarding the fact, that here a model with pure point-spectra
is discussed.

In short, the definition of $S_q$ is based on the following
consideration: If one tries to describe reality in a physical theory, it
is not completely clear with which objects of the theory reality is coupled
\cite{lu}.
The structure of quantum mechanics remains unchanged in the following.
However the concept of a quantum state is looked upon as not immediately
related to physical reality. The relation to it is more intimately given
by the results of measurements on the physical systems being in a certain
quantum-state. The author has heard of this 
approach the first time from W.~Thirring~\cite{thi}.
On the other hand the preference for the concept of a quantum state to
represent the results of measurements (as special
types of events) -- instead of describing an object -- 
is also in a certain concordance with Haag's opinion \cite{haa}
about the meaning of the concept of a quantum-state.

In the spirit of the remarks made before one is led to Shannon's concept
of entropy which was carried further by Kolmogorov and Khinchin
\cite{chi}.
To demonstrate this, one can start with a ``scheme'' (the word used by 
Khinchin),
containing $n$ events. In a trial exactly one of the events
$a_1, a_2 \cdots, a_n$ can take place with the probabilities $w_1, w_2,
\cdots$ resp. $w_n$ and with $\sum w_i$ = 1:
The ``scheme'' is written as
\begin{equation}
 {a_1 a_2 \cdots a_n \choose w_1 w_2 \cdots w_n}\quad  {\rm with}
\quad \sum w_i = 1
\end{equation}
One defines the entropy of such a scheme -- one can also call this
an $n$-alternative with a probability measure -- by the well known expression
\begin{equation}
S = -\sum w_i \ln w_i
\end{equation}
The maximum of $S$ is assumed with $w_i = \frac{1}{n}$ and has the value
$S = \ln n$. The minimum $S$ = 0 is reached, if one of the events say $a_k$
takes place with certainty $w_k$ = 1; the scheme then becomes trivial.

The concept of entropy of that kind can be used in two ways:
\begin{enumerate}
\item
Given a scheme the above defined entropy  is an uncertainty
in the answer for the question, which event actually will happen, when
the scheme will be realized in an experiment.
\item
After the outcome of the experiment is known and in this way the
uncertainty is replaced by certainty the amount of information gained 
can also be expressed by the defined entropy.
\end{enumerate}

In the quantum-mechanical case alternatives with a probability measure
originate for instance in connection with the measurements of observables.
The connection appears most clearly in cases where the observables can be
represented by selfadjoint operators with pure point-spectra. Without
special assumptions one cannot expect that such interesting quantities
for instance as the momentum or the energy of a quantum system have as their
representatives self-adjoint operators with pure point-spectra. On the
other hand, there are known ways -- partly of 
mathematical nature, partly
by somewhat changing the physical situation -- to replace the 
continua in the spectra by point-spectra.
In this case -- assuming in addition that the point-spectrum is 
simple,i.e.\  to each eigenvalue corresponds a one-dimensional
subspace, 
the corresponding scheme for the quantum state $\varphi$ and the observable
$A$ describes the situation:
\begin{equation}
 {a_1, a_2, \cdots, a_k, \cdots \choose w_1, w_2, \cdots w_k, \cdots}
\end{equation}
\noindent
The interpretation for this scheme is:
\begin{equation}
\varphi = \sum c_i \varphi_i, \{A\varphi_i = a_i \varphi_i\}, \{w_i =
\left|c_i\right|^2\}
\end{equation}
\noindent
The corresponding entropy $\tilde{S}(\varphi, A)$ is then
\begin{equation}
\tilde{S}(\varphi, A) = -\sum w_i \ln w_i   \label{1.1}
\end{equation}
\noindent
$\tilde{S}$ is the expression for the uncertainty in the result measuring the
observable $A$ on the system in the state $\varphi $ before the
measurement is made.

Section 2 contains some simple remarks concerning some relations between different
$\tilde{S}(\varphi, A_j)$
by fixing $\varphi $ and varying $A_j$.

The concept of alternatives with probabilities seems to be an
adequate tool for the definition of an entropy like quantity, if one regards
events as the primary concepts linking the theory to the physical reality.
On the other hand one should clarify the relationship between quantities
like $\tilde{S}(\varphi, A_j)$ and the entropy $S$ (defined in the usual
way)
of a system which is based only on the uncertainty concerning the quantum
states. $\tilde{S}(\varphi, A_j)$ is quantitativly equal to the entropy of the
mixture which originates from the measurement of $A_j$. But what does this
have
in common with the usual $S(\varphi)$, which in our case, treated first
for simplicity for $\varphi \in H$,
vanishes? In section 3 we intend by using expressions like
$\tilde{S}(\varphi, A_j)$ to come back to an entropy-like quantity
$S(\varphi)$. However, there remains a conceptional difference between
$S(\varphi)$ and $S_q(\varphi)$. $S(\varphi)$ is based on the more
ontological concept of a state, while $S_q(\varphi)$ expresses 
the knowledge concerning the probabilities of possible events
concentrated in the concept of a state. Therefore $S(\varphi)$ and
$S_q(\varphi)$ may differ from each other.

In section 4 one finds some remarks concerning $S_q$ for particle systems
and in section 5 especially for 2-particle systems. In section 6 a critical
discussion and some concluding remarks are added.

\section
  [Alternatives weighted with Probability Measures (Schemes)
    and their Entropies]
  {Alternatives weighted with Probability\\ Measures (Schemes) 
    and their Entropies}
\setcounter{equation}{0}

Two schemes $M_1$ and $M_2$ may be given. 
Certain conditions may exist between the probabilities of the
events of $M_1$ and those of $M_2$. We are interested
in the simplest situations, where one scheme is finer (or coarser) than
the other. Let's assume $M_1$ is finer than $M_2$. For the members of $M_1$ we
use double indices:
\begin{equation}
M_1: {a_{11}, a_{12}, \cdots, a_{1r}, a_{21}, a_{22}, \cdots a_{2s}, \cdots,
       a_{m1}, a_{m2} \cdots, a_{mz}  \choose
     w_{11}, w_{12}, \cdots, w_{1r}, w_{21}, w_{22}, \cdots w_{2s}, \cdots,
       w_{m1}, w_{m2} \cdots, w_{mz}}
\end{equation}
\begin{equation}
M_2: {b_1, b_2, \cdots, b_m      \choose
     w_1, w_2, \cdots, w_m}
\end{equation}
If one of the events in scheme $M_1$, namely $a_{11}$ or $a_{12}$ or $a_{1r}$
happens, then in $M_2$ event $b_1$ takes place, if $a_{21}$ or $a_{22}$ or
$a_{2s}$ in scheme $M_1$ happens, then $b_2$ in $M_2, \cdots $ and so on.
For the probabilities one assumes
\begin{equation}
\sum_k w_{rk} = w_r, \quad \sum w_r = 1, \quad r = 1,2, \cdots, m
\end{equation}
In this case, where $M_1$ is finer than $M_2$, one can write
\begin{equation}
M_1 \stackrel{f}{\supset}M_2
\end{equation}
It is clear, how one can use the property of being finer or coarser for
observables especially here for those with pure point spectra. Besides
the common domain of definition the corresponding selfadjoint operators must
commute and the property of being finer and coarser should be independent
of the state to which they are applied. If these conditions are
fulfilled, one can introduce sequences of self adjoint operators
\begin{equation}
\label{ops}
A_1 \stackrel{f}{\subset} A_2 \stackrel{f}{\subset} \cdots
\stackrel{f}{\subset} A_n  \label{2,1}
\end{equation}
\noindent
where the operators
become finer going from left to right.
In principle they can end to the right with a selfadjoint operator
with a simple point-spectrum. As one can see easily, the corresponding
entropies increase also in the direction to the right
\begin{equation}
\tilde{S}(\varphi, A_1) \le \tilde{S}(\varphi, A_2) \leq  \cdots
\tilde{S}(\varphi,A_n)
   \label{2,2}
\end{equation}
If as assumed, $\varphi $ is a pure state then $\tilde{S}(\varphi,1)$ = 0,
where 1 is the unit-operator.

That such a sequence becomes arbitrarily 
fine and ends with a selfadjoint operator with a simple 
point-spectrum can be easily fulfilled in a
separable Hilbert space. In the following sections it is however intended to
restrict the observables to a set which is represented by a proper subset of
the set of self-adjoint operators. Therefore the following assumption is not
trivial and generically an idealisation.

{\bf Assumption (F):} Each observable represented by a selfadjoint operator
with a pure point spectrum is an element of a sequence of observables which
ends -- in the direction to become finer -- with an observable having a simple
point spectrum.

To sum up we arrive at the following situation:
\begin{enumerate}
\item
The interesting observables from a physical standpoint
are approximated
by observables, whose representing selfadjoint operators have pure
point-spectra.
\item
Further one assumes that in this approximating set one can find
self-adjoint operators to fulfill (F).
\end{enumerate}

\noindent
Two remarks should be helpful
\begin{enumerate}
\item
An important motivation of this paper is the fact that not all selfadjoint
operators (resp.\ those with pure point spectra) are actual observables.
\item
Generically one can derive other observables
from the spectrum of some observable or from
the spectra of different observables -- if the corresponding selfadjoint
operators commute -- by using real functions of the
values contained in their spectra . For instance for a system with several
particles the total momentum or the total energy is calculated from values
of the momenta and energies measured for each particle separately. This
procedure must be justified by the assumption that the S-operator exists
and therefore the dynamical interactions between the particles can be
neglected. So without the intention to discriminate such observables as
the total momentum, total energy in the case of the particle systems 
we have in
mind the special observables belonging to the measurements made at each
particle. When we refer in the following to the set of observables in the
particle picture we restrict this set in such a way that the measurements
of the physical quantities of each particle are represented separately.
To give a name to this set one can call it: ``Set of actual observables in the
particle picture''. To give a general description of the set of actual
observables, one can remark that they are those which actually transform
by their measurement generically a pure state into a mixture or a
mixture into a more refined mixture.%
\footnote{The remark 2.
          has its origin in a discussion with H.~Roos
          who critisized the too narrow concept of observables in an earlier
          version of this paper}
\end{enumerate}

\section{Definition of the Entropy-like Quantity $S_q$ for Quantum-Mechanical
Systems}
\setcounter{equation}{0}

Although we will use the concept of the entropy-like quantity $S_q$ only
for quantum systems consisting of several particles, we  will describe the
general situation, in which $S_q$ may differ from $S$ for quantum states.

As mentioned before, the concept of a quantum-state will here not be looked upon
as describing an ontological object but as a tool to describe all possible
events in itself or originating from interaction with other systems and
specially to make propositions concerning the probabilities of the results
of measurements in the future.

For the definition of $S_q(\varphi)$ we intend therefore to make use of the
quantities
$\tilde{S}(\varphi, A_j)$
with variable $A_j$. 
One can differentiate between  two properties of the quantities
$\tilde{S}(\varphi, A_j)$.

\begin{enumerate}
\item
If one performs a preparation of a state $\varphi$ as an individual
of an ensemble, then this procedure is intended to create a uniform
ensemble with all individuals in the same state, or in the
language of statistics, to create an ensemble with the highest possible
order. In a certain idealisation the outcome of the procedure will then
be a pure state $\varphi \in H$. Generically the measurement of an
observable is connected with a disturbance of this order (expressed by the
transformation of the pure state $\varphi$ to a mixture). 
In order to define $S_q(\varphi)$ one is inclined to
use those $A_j$ in
$\tilde{S}(\varphi, A_j)$ which lead to the smallest disturbance of this
order.

\item
If on the other hand one regards a sequence of observables as in (\ref{ops})
in the direction to the left, where the observables become coarser, one could
for instance end the procedure with the unit-operator. The effect is then:
No disturbance, however also no information. In the whole, one has the feature
that coarser operators lead to smaller entropies
$\tilde{S}(\varphi,A_j)$.
However, one is interested in gaining as much information about the state as 
possible at all.
\end{enumerate}

Combining 1. and 2. in pursuing the intention to define $S_q(\varphi)$ by quantities
$\tilde{S}(\varphi, A_j)$
one has to look for those $A_j$ leading to the highest possible information,
together with the least disturbance of the quantum state in question.

From remark (2), it follows that only observables $A_j$ with the finest properties
are used for the definition of $S_q(\varphi)$ by the expressions
$\tilde{S}(\varphi, A_j)$.
In this context we use the assumption (F) made in  section 2. 
Let us use the notation ${\bf A}_e$ for the set of actual observables with
simple point-spectra. 
\vskip2mm

\begin{Def}{}
$S_q(\varphi) = \inf\limits_{A_k \in {\bf A}_e}\tilde{S}(\varphi, A_k)$.
\end{Def}{}

\begin{Rem}{1} Let us denote the set of selfadjoint operators with simple
point spectrum  by $\hat{\bf A}_e$. In the case that
${\bf A}_e = \hat{\bf A}_e$ for the quantum-mechanical
system in question, one gets $S_q(\varphi) = 0$. We assumed that
$\varphi $ is a pure state and we can find selfadjoint-operators 
in $\hat{\bf A}_e$  with
simple point-spectra for which $\varphi$ is an eigenstate. 
In these cases -- even
if one includes mixtures besides the pure states -- the $S_q$ defined
above is equal to entropy $S$ as usually defined.
\end{Rem}{}
\begin{Rem}{2} The sets ${\bf A}_e$ resp. $\hat{\bf A}_e$ can be replaced
by the sets ${\bf P}_e$ resp. $\hat{\bf P}_e$
of the minimal projection-operators belonging to the spectral decompositions
of the selfadjoint operators in ${\bf A}_e$ resp $\hat{\bf A}_e$.

The main point of this section is clear from the foregoing: To define $S_q$
one
has to define the set ${\bf A}_e$ of all simple actual observables as a
subset
simple selfadjoint operators $\hat{\bf A}_e$
(with pure point-spectra). If ${\bf A}_e = \hat{\bf A}_e$ is a
good idealisation for
the states of a Hilbert-space {\bf H}
(with a certain physical interpretation),
then $S_q(\phi)$ is the same quantity as $S(\phi)$, where $\phi $ is a pure
sta
or a mixture. In the next section we consider a different physical
situation.
\end{Rem}{}

\begin{Rem}{3} The definition of $S_q(\varphi)$ is based on 
$\tilde{S}(\varphi, A_k)$
with $A_k \in {\bf A}_e$ and also on condition (F) 
which clearly has its origin in the model used.
Physical considerations make it desirable to have selfadjoint observables
with the highest distinguishing power in $\tilde{S}(\varphi, A_n)$.
Therefore one can think for instance of introducing
maximal Abelian subalgebras of selfadjoint operators instead of ${\bf A}_e$.
This idea was also discussed with other
physicists. However, the degree of fineness of the operators involved in the
definition of $S_q$ -- on which for instance the amount of entropy creation
in a
scattering process depends -- has its limitations in physical circumstances
and
is not so much dependent on the mathematical tools one uses.
\end{Rem}{}

\section
{Actual observables for a physical system of $n$-particles}
\setcounter{equation}{0}

The essential concept expressed by the definition of the entropy-like quantity
$S_q$ shall be illustrated with the example of a system consisting of $n$ particles.
The Hilbert space ${\bf H}$ of the total system is constructed from the 1-particle
Hilbert spaces $H^{(k)}$ with $k = 1,2,\cdots n$ by taking the $n$-fold
tensor product and performing after that the completion. If the particles
are of the same kind one has to restrict ${\bf H}$ by symmetry (or antisymmetry)
conditions applied to the combinations of the products of the 1-particle states.
For the physical conditions we have in mind that

\begin{enumerate}
\item
The energy of the particles is low enough to exclude particle creation.
\item
The particles behave like free particles apart from the short time when they are
scattering off each other.
\end{enumerate}

So one can use the mathematical construction of the tensor product of 
1-particle states as a description of the physical situation of the
$n$-particle state in a good approximation.

It is clear what physical conditions are necessary for that: The
$n$-particle state has to be
dilute enough that the time of scattering should be short compared with the time
when the particles move approximately as free particles. This is a stronger
assumption than that of the existence of the S-matrix.

Under these conditions one is able to define the set of observables in a
plausible way in view of the physics involved.

The set of the actual observables is a subset of the set
\begin{equation}
\label{tens}
{\bf A} = \{A_j^{(1)} \otimes A_k^{(2)} \otimes \cdots A_r^{(n)}\}
\end{equation}
Thereby the observables with different upper indices operate in different
1-particle spaces. To avoid any confusion in the registration of the
measurement results (which result belongs to which particle?) or to be
free from correlations arising from the interactions of measuring devices
among themselves, it might sometimes be necessary to restrict the regions
in which the observables in (\ref{tens}) actually operate. However, this
shall 
not be discussed here in detail; only the meaning of the word subset of
the set (\ref{tens}) should be made plausible.

The set ${\bf A}$ of (\ref{tens}) representing the 
actual observables belongs only to a
small subset of the selfadjoint operators  operating in ${\bf H}$.
The observables of (\ref{tens}) have a characteristic property: They transform
superpositions of factorizing states, which usually are the outcome of scattering
processes into the corresponding mixtures, if they have been measured.

As it follows from the content of section 3, 
one has to use proper subsets of the 1-particle operators denoted by
$\{A_j^{(l)}\}$ $l = 1,2, \cdots n$
in (\ref{tens}) 
for the definition of
$S_q(\Phi)$, $\Phi \in {\bf H}$. 
The subsets are collections of those observables
which are as fine as possible -- in the model used here they are represented
by selfadjoint operators with a simple (or nondegenerate) point-spectrum.
If one uses the symbolic notation ${\bf A}_e$ for the set
${\bf A}_e^{(1)}\otimes {\bf A}_e^{(2)}\otimes \cdots {\bf A}_e^{(n)}$
the definition for $S_q$ is then
\begin{equation}
S_q(\Phi) = \inf\limits_{\Phi \in {\bf H},\,C_j \in {\bf A}_e}
  \tilde{S}(\Phi, C_j)\;.  
 \label{4.2}
\end{equation}
The measurements represented by an element of ${\bf A}_e$ (this means by the
tensor product of $n$ self adjoint operators operating in the $n$ 1-particle
spaces) tranforms $\Phi \in {\bf H}$ into the following mixture:
\begin{equation}
\label{sum}
\Phi = \sum_{j_1,j_2, \cdots j_n} \varphi_{j_1} \otimes \cdots \varphi_{j_n}
 \longrightarrow \{\left |c_{j_1,j_2, \cdots j_n} \right |^2
 \varphi_{j_1} \otimes \varphi_{j_2} \otimes \cdots \varphi_{j_n}\}
\end{equation}
\noindent with
\begin{equation}
\sum_{j_1,j_2 \cdots j_n} \left |c_{j_1,j_2,\cdots j_n}\right |^2 = 1
\end{equation}
\noindent if $\|\Phi \|$ = 1.
Thereby the orthonormal basis $\{\varphi_{j_1}\}, \{\varphi_{j_2}\},
\cdots \{\varphi_{j_n}\}$ chosen in each 1-particle space is the set of eigenstates
of the corresponding selfadjoint operators:
\begin{equation}
A\in {\bf A}_e, A = A^{(1)} \otimes A^{(2)} \cdots \otimes A^{(n)},
\{A^{(l)}\varphi_{l_k} = a_{l_k}\varphi_{l_k}\}
\end{equation}
\noindent
 with $l = 1,2 \cdots , n$.

Obviously one gets
\begin{equation}
\tilde{S}(\Phi,A) = -\sum_{j_1,j_2,\cdots j_n}
\left |c_{j_1,j_2, \cdots j_n}\right |^2 \ln
\left |c_{j_1,j_2, \cdots j_n}\right |^2
\end{equation}
\noindent
and one should vary $A \in {\bf A}_e$, defined above to obtain
\begin{equation}
S_q(\Phi) = \inf\limits_{A_j \in {\bf A}_e} \tilde{S}(\Phi,A_j)\;.
\end{equation}
It is a trivial remark that $S(\Phi) = 0$ is valid independently of the kind
of scattering or whether there is any scattering at all.

However by restricting the actual observables by (\ref{tens}) and basing the
entropy-like quantity $S_q$ on the expressions $\tilde{S}(\Phi,A_j)$ one
gets another picture for $S_q$. Considering a single scattering process
described by $\Phi (t)$ (the mathematical details of the asymptotic conditions
should not matter in the moment), and looking to $S_q(\Phi (t))$ at a time
$t_{in}$ before and a time $t_{\rm out}$ after the 
scattering took place one has the following generic picture:

Before the scattering of 2 particles one prepares each particle if possible as a
pure 1-particle state (or at least as a well defined mixture of 1-particle states).
So one has before scattering a state $\varphi^{(1)} \otimes \chi^{(2)}$ 
as a tensor product
(or, taking statistics into account, a symmetrized or antisymmetrized 
tensor product if one has particles of the same kind).

Denoting $\Phi_{in} = \varphi_{in}^{(1)} \otimes \chi_{in}^{(2)}$
the state before the scattering
one can find $A^{(1)}$ resp. $B^{(2)}$ in for which
$\varphi_{in}^{(1)}$ resp.
$\chi_{in}^{(2)}$ are eigenstates of $A^{(1)}$ resp. $B^{(2)}$.

Therefore $S_q(\Phi_{in}) = {\rm inf} \tilde{S}(\Phi_{in}, A_j) = 0$
with $A_j \in {\bf A}_e$.
However, after the scattering process the state
$\Phi_{\rm out}$ does not factorize. Therefore it is not possible to find an
operator $B\in {\bf A}_e$, for which $\Phi_{\rm out}$ is an eigenstate;
this has the consequence $S_q(\Phi_{\rm out}) > 0$.
The change of a factorizing state into a non-factorizing state by a
scattering process is clearly a fundamental feature for quantum-mechanical
scattering processes. If one would try to describe a non-factorizing state in
the particle picture, which means by 1-particle properties, this would 
not be successful; this can only be done after transforming the quantum-state -- here
the state $\Phi_{\rm out}$ -- into a mixture by performing a 
measurement corresponding
to an operator of the kind defined in (\ref{tens}). 
In this sense the quantities
$\tilde{S}(\Phi, A_j)$ are used as quantized expressions 
for the uncertainty in
the results of measurements before the measurements are performed -- the
uncertainty concerns here the question, which factorizing state would come out
in a measuring process corresponding to (\ref{sum}). 
One can find
some discussion in the concluding remarks of section 6 concerning the 
mathematical frame using in principle $n$ measurements performed 
for some $n$-particle state -- 
leading to a description which does not correspond to the physical situation, if
the number $n$ of particles is large compared to 2.

\section
{Some remarks concerning $S_q$ for 2-particle-states}
\setcounter{equation}{0}

The content of this section is the discussion of the quantities $S_q$ in the case
of 2-particle-quantum-states. We use here the Schr\"odinger picture. It is a
trivial consequence of the discussion in section 3  and section 4 that one is
not able to define $S_q$ as a characteristic quantity for a Heisenberg state.

The state $\Phi(t_{in}) = \Phi_{in}$ before the scattering is transformed by the
scattering process into $\Phi(t_{\rm out}) = \Phi_{\rm out}$. 
For the considerations here
it is not important whether one is able to describe the scattering by a unitary
time development $\Phi(t) = U(t,t_{in})\Phi_{in}$ coming to
$\Phi_{\rm out} = U(t_{\rm out},t_{in})\Phi_{in}$ or to come from $\Phi_{in}$ immediately
to $\Phi_{\rm out}$ by applying to $\Phi_{in}$ a scattering operator
(S-matrix).
In a certain idealization the preparation of $\Phi_{in}$ might lead to
\begin{equation}
\label{prod}
\Phi_{in} = \psi_{in}^{(1)} \otimes \chi_{in}^{(2)},
\end{equation}
\noindent with
\begin{equation}
\label{dom}
\psi_{in}^{(1)}\in {\bf H}^{(1)}, \chi^{(2)}_{in} \in {\bf H}^{(2)},
\|\psi_{in}^{(1)} \| = 1,
\|\chi_{in}^{(2)} \| = 1.
\end{equation}
\noindent
(For reasons of simplicity it is assumed here 
that the particles are of 
different type, so it is not necessary to symmetrize or
antisymmetrize.)
One sees that for
\begin{equation}
\label{seq}
S_q(\Phi_{in}) = inf \tilde{S}(\Phi_{in}, A_j \otimes B_k), 
\quad A_j\in {\bf A}_e^{(1)},
B_k\in {\bf A}_e^{(2)}
\end{equation}
\begin{equation}
S_q(\Phi_{in}) = 0
\end{equation}
The property that $S_q(\Phi_{in})$ vanishes is based on the fact that one can
find in ${\bf A}_e^{(1)}$  as well as 
in  ${\bf A}_e^{(2)}$  observables, which are represented by 
self-adjoint operators
with simple point-spectra and for which $\psi_{in}^{(1)}$ resp.
$\psi_{in}^{(2)}$ are eigenstates.

\noindent
$\Phi_{\rm out}$ can be described by
\begin{equation}
\Phi_{\rm out} = \sum_{i,k}c_{ik} \hat{\varphi}_i \otimes \hat{\eta}_k,
\quad \sum_{i,k} \left |c_{ik}\right |^2 = 1
\end{equation}
\noindent
with $\{\hat{\varphi}_i\}$ resp. $\{\hat{\varphi}_k\}$
as orthonormal systems in
${\bf H}^{(1)}$ resp. ${\bf H}^{(2)}$.

J.~v.~Neumann has shown that one can find orthonormal
systems $\{\varphi_l\}$ resp. $\{\eta_k\}$ (dependent on $\Phi$), 
to bring $\Phi $, here in particular
$\Phi_{\rm out} $, into the normal form
\begin{equation}
\label{sumfi}
\Phi_{\rm out} = \sum_e \sqrt{w_l} \varphi_l \otimes \eta_l
\end{equation}
\noindent
with $\sum w_l = 1$.

If in the sum of (\ref{sumfi}) some $w_i$ coincide, one gets a variety of
normal forms for one and the same state. If for instance for a certain
state $\Phi$ one has a total degeneration of the ${w_i}$, one gets
\begin{equation}
\Phi = \frac{1}{\sqrt{n}} \sum^n_{l=1} \varphi_l \otimes \eta_l = \frac{1}{\sqrt{n}}
     \sum ^n_{l=1} \varphi'_l \otimes \eta'_{l}
\end{equation}
\noindent   with
\begin{equation}
\varphi_l = \sum^n_{s=1} u_{sl}\varphi'_s, \quad \eta_l = \sum ^n_{s=1}
\bar u_{sl} \eta'_s
\end{equation}
Thereby $\{u_{sl}\}$ is an arbitrary unitary $n\times n$-matrix and 
$\{\bar u_{sl}\}$
is its complex-conjugate.
The normal forms of the state $\Phi_{\rm out} $ show in their structure the correlations
originating from the conservation of physical quantities like total momentum or
total energy. The point spectrum would physically correspond to the 
enclosure of the
2-particle system in a box. (Such an idealization comes, however, somewhat in
conflict with the S-matrix picture.)

On the other hand the normal forms give a hint which pairs of observables 
one should
use as factors in the tensor product $A_f \otimes B_g$ to obtain 
the infimum, which here
becomes the minimum. Before demonstrating this, it is useful to 
give a 
\vskip2mm
\begin{Def}{}

A state $\Phi \in {\bf H}$, for instance $\Phi_{\rm out}$ in (\ref{prod}),
may be represented in a normal form
\begin{equation}
\Phi = \sum \sqrt{w_l} \varphi_l \otimes \eta_l
\end{equation}
As a second mathematical object there should be given
a pair of self-adjoint operators with
pure simple point-spectra and their tensor-product $A \otimes B$.
We say $A \otimes B$ is
adapted to the given normal form of $\Phi $ if $\{\varphi_l\}$
resp. $\{\eta_l\}$ is
in the set of eigenvectors of $A$ resp. $B$.

\end{Def}{}
With this definition one is able to formulate the
\vskip2mm
\begin{Prop}{}
The quantity
\begin{equation}
\inf\limits_{C_j\in{\bf A}^{(1)}_e,\, D_k \in {\bf A}^{(2)}_e}
\tilde{S}(\Phi, C_j \otimes D_k)
\end{equation}
is reached by a pair $A \otimes B$,
which is adopted to one and the same normal form of $\Phi $.
\end{Prop}{}

\begin{Rem}{1}
 The infimum is actually a minimum and one gets
\begin{equation}
S_q(\Phi) = {\rm min} \tilde{S}(\Phi, C_j \otimes D_k) = 
\tilde{S}(\Phi, A \otimes B) =
-\sum_l w_l \ln w_l
\end{equation}
\end{Rem}{}
\begin{Rem}{2}
 If there exist several normal forms for $\Phi $ and the
tensor product $A' \otimes B'$, with $A'\in {\bf A}_e^{(1)}$,
$B' \in {\bf A}_e^{(2)}$ is
adapted to another normal form of $\Phi $, then again one gets
\begin{equation}
S_q(\Phi) = \tilde{S}(\Phi, A' \otimes B') = -\sum_l w_l \ln w_l
\end{equation}
\end{Rem}{}
\begin{Prf} We assume that both factors in $A \otimes B$ are 
adapted to the same normal 
form
of $\Phi $. $C \otimes D$ should be constructed with arbitrary 
operators $C$ and
$D$, fulfilling the conditions $A, C \in {\bf A}_e^{(1)} and
B, D \in {\bf A}_e^{(2)}$.
The proposition has been proven if one can show the validity of the
following inequalities resp. equalities
\begin{equation}
\label{equal}
\tilde{S}(\Phi, C \otimes D) \geq \tilde{S}(\Phi, C \otimes 1) \geq
\tilde{S}(\Phi, A \otimes 1) =    \tilde{S}(\Phi, A \otimes B)    
\end{equation}
We prove the different parts of (\ref{equal}):

$\tilde{S}(\Phi, C \otimes D) \geq
\tilde{S}(\Phi, C \otimes 1)$ holds, because $C \otimes D$
is finer than $C \otimes 1$. In addition
$\tilde{S}(\Phi, A \otimes 1) = \tilde{S}(\Phi, A \otimes B)$, 
since $A \otimes B$ is adapted
to one and the same normal form of $\Phi $. With $\Phi = \sum\sqrt{w_l}
\varphi_l \otimes \eta_l$ this is true, since
\begin{equation}
(P_k \otimes 1)\Phi = (P_k \otimes Q_k)\Phi =
(1 \otimes Q_k)\Phi = \sqrt{w_k}\varphi_k \otimes \eta_k
\end{equation}
\noindent
can be derived from the properties
$ P_k\varphi_l = \delta_{kl} \varphi_l,\ Q_k\eta_l = \delta_{kl}\eta_l. $

\noindent
The remaining task is to prove
\begin{equation}
\tilde{S}(\Phi, C \otimes 1) \geq \tilde{S}(\Phi,A\otimes1) = -\sum w_l\ln w_l
\end{equation}
\noindent
Let now $\{\hat{\varphi}_s\}$ be an orthonormal basis of
eigenvectors of $C$.
\noindent
We use the substitution
\begin{equation}
\{\varphi_l = \sum_su_{sl}\hat{\varphi}_s\}
\end{equation}
\noindent
with $\{u_{s_l}\}$ as unitary matrix. Introducing this substitution into
(\ref{dom}) one gets
\begin{equation}
\Phi = \sum_l\sqrt{w_l} \varphi_l \otimes \eta_l = \sum_l\sqrt{w_l}\sum_su_{sl}
   \hat{\varphi}_s \otimes \eta_l
\end{equation}
\noindent
and with $\hat P_s \hat{\varphi}_t = \delta_{st} \hat{\varphi}_t$
\begin{eqnarray}
\|\hat P_s \otimes 1\Phi \|^2 &=& (\Phi,\hat P_s \otimes 1\Phi)
\nonumber\\&=&
\left(\sum_l\sqrt{w_l}\sum_t u_{tl}\hat {\varphi}_t 
\otimes \eta_l, \sum_m\sqrt{w_m}
 u_{sm}\hat{\varphi}_s \otimes \eta_m\right)
\\&=&
\sum_lw_l \left |u_{sl}\right |^2
\nonumber
\end{eqnarray}
So it follows
\begin{equation}
\tilde{S}(\Phi, C \otimes 1) = -\sum_s\sum_lw_l\left|u_{sl}\right |^2 \ln
\sum_mw_m \left |u_{sm} \right |^2.
\end{equation}

\noindent
Since $f(x)=x\ln x$ is a convex function
\begin{equation}
f(\sum_lp_lx_l) \leq \sum_lp_lf(x_l)
\end{equation}
provided
\begin{equation}
 0 \leq p_l \leq 1, l = 1, 2, \cdots, \sum p_l = 1.
\end{equation}
\noindent
Introducing $\left |u_{sl}\right |^2 = p_l,\ w_l = x_l$
the inequality
\begin{equation}
\sum_lw_l\left |u_{sl}\right |^2\ln \sum_mw_m\left |u_{sm}\right |^2
\leq \sum_l \left | u_{sl} \right |^2 w_l\ln w_l
\end{equation}
\noindent
holds for each $s$. Therefore one gets
\begin{equation}
\sum_s\sum_lw_l\left |u_{sl}\right |^2 \ln \sum_mw_m\left |u_{sm}\right |^2
\leq \sum_lw_l\ln w_l
\end{equation}
\noindent  and from (5.7)
\begin{equation}
\tilde{S}(\Phi, C \otimes 1) \geq -\sum_lw_l \ln w_l = \tilde{S}(\Phi, A \otimes B)
\end{equation}
\noindent
This completes the proof.
\end{Prf}
One may ask the question in which situation the equation
\begin{equation}
\tilde{S}(\Phi, A \otimes 1) = \tilde{S}(\Phi, C \otimes 1)
\end{equation}
\noindent holds.

We discuss at first the case when all $w_k$ are different from each
other. There is a possibility that in the
substitution for fixed $s$ a certain $k$ exists, that $\left |u_{sk}\right |^2 = 1$,
while $\left |u_{sl}\right |^2 = 0$ for $l \neq k$. If this same property is true
for every $s$, then the equation $\tilde{S}(\Phi, A \otimes 1) = \tilde{S}(\Phi, C \otimes 1)$
is valid. On the other hand the unitary matrix $(u_{sk})$ in the substitution
has only the effect to permute the eigenstates.

A bit more complicated is the case that some (or all) of the $w_k$ are
equal. For the corresponding step in the proof, where the convex function
$f(x)$
is involved, one has then to take this function several times for the same
values of the arguments
by introducing the different values $x = w_l$ into the
convexity condition. Clearly the convexity condition is also 
valid for this case.
On the other hand one is able to arrive at other normal forms for $\Phi $ by using
unitary substitutions in the subspaces with equal values $w_l $.
Then clearly one has a greater variety, besides $A$ also other $C_j \in
{\bf A}_e^{(1)}$, if they leave these subspaces invariant. Then 
$\tilde{S}(\Phi, A \otimes 1) = \tilde{S}(\Phi, C_j \otimes 1)$ follows.
If one takes into account that one wants also 
$\tilde{S}(\Phi, C_j \otimes 1) =
\tilde{S}(\Phi, C_j \otimes D_k)$,
it is clear that again $C_j \otimes D_k$ must be adapted to a 
new normal form, which
one obtains from unitary substitutions characterized above in (\ref{seq}).
This means -- when $C_j$ is constructed by a unitary substitution
$(u_{sl})$ from $A$ as characterized above -- $D_k$ must be obtained by 
the
unitary substitution $(\bar u_{ls})$ from $B$.

\section{Conclusions and some hints for a further approach}

If the actual observables are only a proper subset of the self-adjoint
operators, then the above defined entropy-like quantity $S_q$ differs from
the usual entropy $S$. $S_q(\varphi)$ can be different from zero 
also for pure
states $\varphi $ and $S_q(\varphi)$ can grow with time; this is a
special feature of scattering processes. We showed the mechanism, which
leads to an increase of $S_q$ for the 2-particle-system without using the
details of the scattering. This was demonstrated by using von Neumann's
standard forms for the state after scattering. To make the physical picture
consistent one should show by further considerations that the different
components of the normal form belonging to different factorizing states
can really be observed independently. That is a tedious discussion -- as it
is often the case when physical consideration must be introduced into the
mathematical framework -- which nearly always contains elements of
idealisation.

On the other hand the idealisation introduced above 
for the characterization
of the set of actual observables
${\bf A}_l = A_e^{(1)} \otimes A_e^{(2)} \cdots \otimes A_e^{(n)}$ 
becomes very idealistic,
if $n$ is large and if it is taken seriously as an expression for a measuring
operation. In that case it is natural to use another picture -- 2-particle
scattering states on a background consisting of the $(n-2)$
particle-system --, leading to a steady production of the
entropy-like quantity $S_q$. This is a feature which $S_q$ has in 
common with the
Kolmogorov-Sinai-entropy, although the entropy-production mechanism seems
to be different. I think it is not worthless to study the entropy-production
of $S_q$ for particle systems with large $n$, using the considerations
referring to the 2-particle scattering states.

\section{Acknowledgements}

I want to thank W.~Thirring for a discussion and his hint concerning the
concept of Kolmogorov-Sinai entropy.
Further it is a pleasure and a duty to mention colleagues at the Max Planck
Institute, P.~Breitenlohner, W.~Zimmermann, D.~Maison and E.~Seiler.

\end{document}